\documentclass[superscriptaddress,showpacs,amsmath,amssymb,twocolumn,nofootinbib]{revtex4-1}

\usepackage{amsmath, amsthm, amssymb, bm, braket}
\usepackage[dvips]{graphicx}
\usepackage{color}
\usepackage{wrapfig}
\usepackage{fancybox}

\begin{document}
\title{One-proton emission from the $^6_{\Lambda}$Li hypernucleus}

\author{Tomohiro Oishi}
\email[E-mail: ]{toishi@pd.infn.it} %{gz2r2dhh@gmail.com}
\affiliation{Department of Physics and Astronomy ``Galileo Galilei'', 
University of Padova, and} %via F. Marzolo 8, I-35131, Padova, Italy}
\affiliation{I.N.F.N. Sezione di Padova, \\
via F. Marzolo 8, I-35131, Padova, Italy}

%\author{Lorenzo Fortunato}
%\email[Electronic address: ]{fortunato.lorenzo@pd.infn.it} %{gz2r2dhh@gmail.com}
%\affiliation{Department of Physics and Astronomy ``Galileo Galilei'', 
%University of Padova, and I.N.F.N. Sezione di Padova, 
%via F. Marzolo 8, I-35131, Padova, Italy}

\renewcommand{\figurename}{FIG.}
\renewcommand{\tablename}{TABLE}

\newcommand{\bi}[1]{\ensuremath{\boldsymbol{#1}}}
\newcommand{\unit}[1]{\ensuremath{\mathrm{#1}}}
\newcommand{\oprt}[1]{\ensuremath{\hat{\mathcal{#1}}}}
\newcommand{\abs}[1]{\ensuremath{\left| #1 \right|}}

\def \beq{\begin{equation}}
\def \eeq{\end{equation}}
\def \beqa{\begin{eqnarray}}
\def \eeqa{\end{eqnarray}}
\def \Schr{Schr\"odinger }

\def \bir{\bi{r}}
\def \ubir{\bar{\bi{r}}}
\def \bip{\bi{p}}
\def \ubip{\bar{\bi{r}}}

\begin{abstract}
One-proton (1p) radioactive emission under the influence of 
the $\Lambda^0$-hyperon inclusion is discussed. 
I investigate the hyper-1p emitter, $^6_{\Lambda}$Li, 
with a time-dependent three-body model. 
Two-body interactions for $\alpha$-proton and $\alpha$-$\Lambda^0$ 
subsystems are determined consistently to their resonant and 
bound energies, respectively. 
For a proton-$\Lambda^0$ subsystem, a contact interaction, which 
can be linked to the vacuum-scattering length of the 
proton-$\Lambda^0$ scattering, is employed. 
A noticeable sensitivity of the 1p-emission observables 
to the scattering length of the proton-$\Lambda^0$ 
interaction is shown. 
The $\Lambda^0$-hyperon inclusion leads to a remarkable fall 
of the 1p-resonance energy and width from the hyperon-less 
$\alpha$-proton resonance. 
For some empirical values of the proton-$\Lambda^0$ scattering length, 
the 1p-resonance width is suggested to be of 
the order of $0.1-0.01$ MeV. 
Thus, the 1p emission from $^6_{\Lambda}$Li may occur in the timescale of 
$10^{-20}-10^{-21}$ seconds, which is sufficiently shorter than the 
self-decay lifetime of $\Lambda^0$, $10^{-10}$ seconds. 
By taking the spin-dependence of the 
proton-$\Lambda^0$ interaction into account, 
a remarkable split of the $J^{\pi}=1^-$ and $2^-$ 
1p-resonance states is predicted. 
It is also suggested that, if the spin-singlet proton-$\Lambda^0$ 
interaction is sufficiently attractive, the 1p emission 
from the $1^-$ ground state is forbidden. 
From these results, I conclude that the 1p emission can be a 
suitable phenomenon to investigate the basic properties of 
the hypernuclear interaction, for which a direct 
measurement is still difficult. 
\end{abstract}

\maketitle

%%%%%%%%%%%%%%%%%%%%%%%%%%%%%%%%%%%%%%%%%%%%%%%%%%%%%%%%%%%%%%%%%%%%%%%%%%%
\section{Introduction} \label{sec:intro}
Inclusion of hyperons in atomic nuclei has been a fascinating topic in nuclear physics \cite{2016Gal,2015Feli_Rev,2012Botta_Rev}. %{2006Hashi,2005Tamura}. 
One of the famous effects is that the hyperon plays as a 
gluelike particle: its inclusion leads to an expansion of the proton and neutron driplines, 
as well as a new emergence of stable nuclei. %\cite{1983Motoba,1985Motoba,1996Hiyama,2009Hiyama,2010Harada}. 
Recently, one novel type of the hypernuclear experiment combined with 
the heavy-ion collision has been proposed \cite{2016Rapp,2012Stein,1984Asai,1994Balt}. 
An advantage of this new experimental method is an accessibility 
to the proton-rich and neutron-rich regions, %\cite{2016Rapp,2012Stein}, 
where the inclusion 
effect can provide a noticeable difference from the normal nuclei. 

The fundamental parts of the hypernuclear physics include 
hyperon-nucleon (YN) and hyperon-hyperon (YY) interactions. 
However, even with the modern experimental development, 
a direct measurement of the YN and/or YY 
interactions is still difficult. 
This lack of information has been indeed a long-standing bottleneck to establish 
the unified nuclear model for normal and hyper nuclei. 

On the other hand, in recent decades, there has been a remarkable development 
of experiments to measure the few-nucleon radioactive emissions. 
Those include, e.g. one-proton (1p) and two-proton 
emissions \cite{02Pfu,02Gio,2012Pfu_rev,2009Gri_rev}. %\cite{02Pfu,02Gio,08Blank_01,08Blank_02,2012Pfu_rev,2009Gri_rev}. 
For these radioactive processes, typical observable quantities include 
the few-nucleon resonance energy and its decay width or equivalently lifetime. 
Here it is worth mentioning that the decay width is available 
only in these meta-stable systems, in contrast to 
the bound systems, whose lifetime is trivially infinite. 
It has been expected that these observables can be good reference 
quantities for the theoretical models of the nuclear 
force, pairing correlation, and multi-particle 
dynamics. %\cite{2009Gri_rev,2000Gri_PRL,09Gri_80,09Gri_677,05Flam,07Bertulani_34,12Maru,13Deli,2015Lundmark}. 
Similarly, the particle-unbound hyper nuclei can be a good 
testing field to investigate the hypernuclear properties. %as well as the bound ones. 
There have been, however, still less studies of 
the hyperon-inclusion effect on the particle-unbound 
systems \cite{1983Motoba,1996Hiyama,2008Bely,2009Hiyama,2015Hiyama}. %\cite{1985Motoba,2014Hiyama}. 

%\textcolor{red}{(A) The relevant topics include ab initio 
%calculations, %\cite{02Nemura,03Nemura,2003Usm,2005Nemura,2014Wirth,2016Gazda,2016Wirth}, 
%the realistic YN and YY 
%interactions, %\cite{1994Yamamoto,2005Yamamoto,2010Nemura,2013Nogga,2013Haid}, 
%double-strangeness systems, %\cite{03Nemura,2003Kohno,2003Fili,2005Nemura}, 
%the spin-dependent charge-symmetry breaking, %\cite{2015Yama,2016Gazda}, 
%the tensor-force effect, %\cite{04Ukai,2010Toki}, 
%and the three- or four-body hypernuclear 
%interactions. } %\cite{02Nemura,2003Usm,2016Wirth}. 

The aim of this work is to invoke the interest to utilize the 
proton emission as a suitable tool to investigate the 
hypernuclear properties. 
For this purpose, I demonstrate a simple calculation 
for the lightest hyper-1p emitter, $^6_{\Lambda}$Li. 
The calculation is based on the $\alpha$-proton-$\Lambda^0$ three-body model 
to study the inclusion effect of $\Lambda^0$ hyperon 
on the proton-emitting nucleus. 
For the description of the particle-resonance, or equivalently, 
the meta-stable properties, it is necessary to 
extend the static quantum mechanics. 
The popular methods for this purpose include the time-dependent 
method \cite{04Gurv,14Oishi,17Oishi,2017Ordo}, %\cite{2008Hatano,87Gurv,88Gurv,04Gurv,14Oishi,17Oishi,2017Ordo}, 
as well as the non-Hermitian 
method \cite{1983Ho,2002Michel,14Myo_Rev,2015Hiyama}. %\cite{1983Ho,89Bohm,1996Hatano,2002Michel,06Hagen,14Myo_Rev}. 
In this paper, I employ the former option, 
which can provide one phenomenological 
way to find a useful link between the 
resonance observables and the basic properties in hypernuclear physics. 
Especially, it is expected that the measurement of 
the proton emission provides suitable data to examine 
the YN interaction model. %for which a direct experimental survey is still difficult. 
Note that the similar interest has been devoted to the two-proton 
radioactive emission, where its decay width can be closely 
related to the proton-proton scattering length \cite{2000Gri_PRL,17Oishi}. %07Gri_III,
%Note that the similar time-dependent model was utilized in the previous 
%work on the two-proton emitter, $^6$Be \cite{14Oishi,17Oishi}. 

In the next section, the formalism of the three-body model is presented. 
Parameters of the model are chosen consistently 
with the experimental data. 
In section \ref{sec:3}, the time-dependent calculation for 
the hyper-1p emitter, $^6_{\Lambda}$Li, is presented. 
There, my result and discussion on the hyperon-inclusion 
effect on the 1p emission are described in detail. 
Finally, in section \ref{sec:4}, I summarize this work.

%%%%%%%%%%%%%%%%%%%%%%%%%%%%%%%%%%%%%%%%%%%%%%%%%%%%%%%%%%%%%%%%%%%%%%%%%%%
\section{Three-body Model} \label{sec:2}
For simplicity, I omit the superscript $0$ for the $\Lambda^0$ 
particle in the following. 
The theoretical model is based on the $\alpha$-proton-$\Lambda$ 
three-body picture, within the core-orbital coordinates, 
$\left\{ \bir_{\rm p},\bir_{\Lambda} \right\}$, 
as shown in Fig. \ref{fig:01_3bc}. 
The detailed formalism of these coordinates is summarized in the Appendix. 
In this framework, the total Hamiltonian reads \cite{1991BE,1997EBH,2005HS} 
\beqa
 H_{3b} &=& h_{\rm p} + h_{\Lambda} + x_{\rm rec} + v_{\rm p-\Lambda}(\bir_{\rm p}, \bir_{\Lambda}), \nonumber \\
 h_i &=& \frac{p_i^2}{2\mu_i} + V_{\alpha-i}(r_i), \nonumber \\
 x_{\rm rec} &=& \frac{\bip_{\rm p} \cdot \bip_{\Lambda}}{m_{\rm c}}~~~({\rm recoil~term}), \label{eq:fevag}
\eeqa
where $i={\rm p}$ and $\Lambda$ for the valence 
proton and $\Lambda$ hyperon, respectively. 
Here $\bir_i$ is the relative coordinate between 
the core and the $i$-th nucleon. 
Thus, $h_i$ is the corresponding single particle (s.p.) Hamiltonian. 
Mass parameters are fixed as the empirical values: 
$\mu_i=m_i m_{\rm c}/(m_i + m_{\rm c})$, 
$m_{\rm p}=938.272$ MeV$/c^2$, 
$m_{\Lambda}=1115.683$ MeV$/c^2$, and 
$m_{\rm c} = 3727.379$ MeV$/c^2$ ($\alpha$-particle mass).

\begin{figure}[tb] \begin{center}
 \includegraphics[width=0.75\hsize]{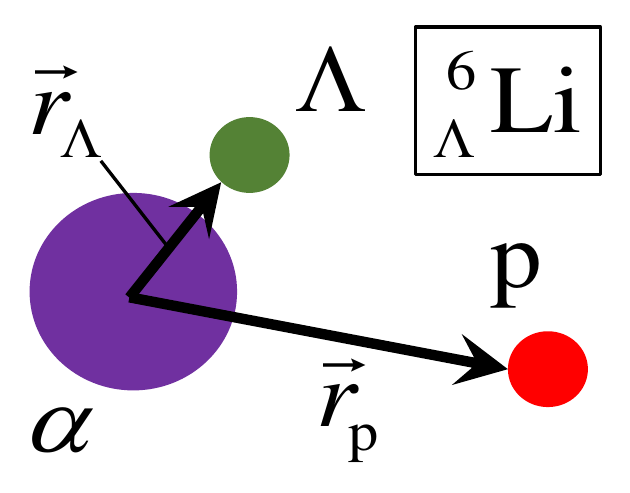}
 \caption{Three-body model of $^{6}_{\Lambda}$Li.} \label{fig:01_3bc}
\end{center} \end{figure}

\begin{table}[b] \begin{center}
\caption{Resonance energy and width obtained with the 
$\alpha$-nucleon potential in this work. 
Those are evaluated from the scattering 
phase-shift in the $(p_{3/2})$-channel. } \label{table:fwure}
  \catcode`? = \active \def?{\phantom{0}} %define `?' as ' '(one-blank).
  \begingroup \renewcommand{\arraystretch}{1.2}
  \begin{tabular*}{\hsize} { @{\extracolsep{\fill}} cccc } \hline \hline
         &   & $E_{\rm \alpha-n},~\Gamma_{\rm \alpha-n}$  & $E_{\rm \alpha-p},~\Gamma_{\rm \alpha-p}$  \\
         &   & (MeV)                      & (MeV)                      \\ \hline
  ~Our $V_{\rm \alpha-n,p}$        &   &$0.77,~0.67$         & $1.69,~1.31$  \\ %Truelly, 1.61 MeV ... %~Ref. \cite{14Oishi}  & $1.032,~0.967$         & $1.954,~1.693$  \\
  ~Exp.\cite{02Till}   &&$0.798,~0.578$   &$1.69,~1.06$ \\
  ~Exp.\cite{88Ajzen}  &&$0.89,~0.60$   &$1.97,~(\simeq 1.5)$ \\
  ~Exp.\cite{NNDCHP}   &&$0.735,~0.60$   &$1.96(5),~(\simeq 1.5)$~ \\ \hline \hline
  \end{tabular*}
  \endgroup
  \catcode`? = 12 %initialize `?'.
\end{center} \end{table}

The core-proton potential is determined as 
\beq
 V_{\rm \alpha-p}(\bir_{\rm p}) = V_{WS}(\bir_{\rm p}) + V_{Coul}(\bir_{\rm p}), \label{eq:mx6}
\eeq
where a Woods-Saxon potential including the spin-orbit term is given as 
\beqa
 V_{WS}(r) &=& V_0 f(r) + U_{ls} (\bi{l} \cdot \bi{s}) \frac{1}{r} \frac{df(r)}{dr}, \label{eq:cp-WS} \\
 f(r) &=& \frac{1}{1 + e^{(r-R_0)/a_0}}. \label{eq:coredens}
\eeqa
Here $r=\abs{\bir}$ and $f(r)$ is a standard Fermi profile. 
In addition, Coulomb potential of an uniformly charged sphere with 
its radius $R_0$ is included for this subsystem. 
In this paper, its parameters are fixed as 
$R_0=r_0\cdot 4^{1/3}$, $r_0=1.25$ fm, $a_0=0.65$ fm, $V_0=-47.4$ MeV, and 
$U_{ls}=-0.4092V_0 r_0^2$ \cite{1997EBH,2005HS}. 
From the phase-shift analysis, it is confirmed that 
this set of parameters fairly reproduces the empirical 
$\alpha$-neutron and $\alpha$-proton scattering data 
in the $p_{3/2}$ channel \cite{02Till}, 
as summarized in TABLE \ref{table:fwure}. 
On the other hand, the core-$\Lambda$ potential is given as 
\beq
 V_{\alpha-\Lambda}(\bir_{\Lambda}) = f V_{WS}(r_{\Lambda}), 
\eeq
with the reduction factor, $f=0.5333$, where the non-physically large 
spin-orbit component in Eq. (\ref{eq:cp-WS}) is not operative. 
This potential correctly reproduces the s.p. binding energy 
in the core-$\Lambda$ subsystem ($^5_{\Lambda}$He) in the $s_{1/2}$ channel, 
$\epsilon(s_{1/2})=-3.12$ MeV \cite{1975Gal}.

\subsection{Proton-$\Lambda$ interaction}
In this work, for the proton-$\Lambda$ subsystem, 
I employ the simple contact-type interaction: 
\beq
 v_{\rm p-\Lambda}(\bir_{\rm p},\bir_{\Lambda}) =  w_0 \delta(\bir_{\rm p}-\bir_{\Lambda}). \label{eq:yutwo}
\eeq
As is well known, its bare strength, $w_0$, can be determined 
consistently to the vacuum proton-$\Lambda$ scattering length, $a_v$, 
by determining also the energy cutoff $E_{\rm cut}$ 
of the model space \cite{1991BE,1997EBH}. 
That is, 
\beq
 a_v = \left[ \frac{2k_{\rm cut}}{\pi} 
 + \frac{2\pi \hbar^2}{\mu_{\rm p-\Lambda} w_0} \right]^{-1}~~({\rm fm}), \label{eq:orocs}
\eeq
or equivalently, 
\beq
 w_0 = \frac{2\pi^2 \hbar^2 a_v}{\mu_{\rm p-\Lambda} (\pi-2a_v k_{\rm cut})} 
~~~~({\rm MeVfm^3}), 
\eeq
where $\mu_{\rm p-\Lambda}=m_{\rm p} m_{\Lambda}/(m_{\rm p}+m_{\Lambda})$ and 
$k_{\rm cut}=\sqrt{2\mu_{\rm p-\Lambda} E_{\rm cut}}/\hbar$. 
The energy cutoff is fixed as $E_{\rm cut}=15$ MeV in this paper. 
Unfortunately, there has been no direct measurement available 
for the $\Lambda-p$ scattering length in vacuum. 
Thus, in the following discussion, I treat it as a model parameter, 
and results with several $a_v$ values are compared. 
Also, in Sec. \ref{subsec:sd_split}, the model is extended to 
take the spin-dependence of the scattering length into account.

%\begin{figure}[tb] \begin{center}
% \includegraphics[width=\hsize]{hoge_01_wa.pdf}
% \caption{Scattering length, $a_v$, of the proton-$\Lambda$ scattering 
%with the contact interaction strength, $w_0$.} \label{fig:02_wa}
%\end{center} \end{figure}

%Figure \ref{fig:02_wa} shows the relationship given in Eq. (\ref{eq:orocs}). 
%Until the bound-unbound border, $w_0 \simeq -1200$ MeVfm$^3$ in our case, 
%the stronger attraction corresponds to the lower $a_v$ value. 
%Notice also that, for $w_0$ lower than the bound-unbound border, 
%the proton-$\Lambda$ subsystem fictionally becomes bound, 
%and the scattering length unphysically gets positive. 

Note that there are several, more realistic YN-interaction models than 
the contact type \cite{2016Gal,2015Feli_Rev}. 
However, these interactions have finite ranges. 
At present, the time-dependent calculation combined with the finite-range 
proton-$\Lambda$ force is not feasible, due to the numerical cost. 
The main problem is that, for $J^{\pi}=1^-$ or $2^-$, one needs a larger 
model space than in the case of $J^{\pi}=0^+$ \cite{14Oishi,17Oishi}. 
To employ these finite-range forces, the algorithmic improvement 
as well as the parallelized computation is in progress now.

\subsection{Uncorrelated basis}
In this work, I assume that the $\alpha$ core keeps 
the spin-parity of $0^+$ without excitation. 
Then, the proton-$\Lambda$ state is expanded on the uncorrelated basis, 
which is the tensor product of two s.p. states coupled to $J^{\pi}$. 
That is, 
\beq
 \ket{\Phi^{({\rm p\Lambda},J^{\pi})}_{\kappa \lambda} }
 = \left[ \ket{\phi^{\rm (p)}_{\kappa}} \otimes \ket{\phi^{(\Lambda)}_{\lambda}} \right]^{(J^{\pi})}, 
\eeq
where $\kappa$ is the shorthanded label for all the quantum numbers 
of the proton state, $\left\{ n_p,l_p,j_p,m_p \right\}$, and 
similarly with $\lambda$ for the $\Lambda$ state. 
Those include the radial quantum number $n$, 
the orbital angular momentum $l$, 
the spin-coupled angular momentum $j$, and 
the magnetic quantum number $m$. 
Of course, each s.p. state satisfies, 
\beq 
 h_{\rm p} \ket{\phi^{\rm (p)}_{\kappa}} = \epsilon_{\kappa } \ket{\phi^{\rm (p)}_{\kappa}},~~~
 h_{\Lambda} \ket{\phi^{(\Lambda)}_{\lambda}} = \epsilon_{\lambda} \ket{\phi^{(\Lambda)}_{\lambda}},
\eeq 
where $\epsilon_{\kappa(\lambda)}$ is the single-proton ($\Lambda$) energy. 
In order to take into account the Pauli principle for the valence proton, 
the first $(s_{1/2})$ state is excluded. 
The s.p. angular momenta up to the $l_{\rm cut}=3$ are taken into account. 
The continuum s.p. states up to $E_{\rm cut}=15$ MeV 
are discretized in the radial box of $R_{\rm box}=120$ fm. 
This truncation of model space is confirmed to be in a range that ensures 
the convergence in the 1p-resonance observables.

The three-body eigenstates are solved by diagonalizing 
the Hamiltonian matrix. %Those states can be expanded on the uncorrelated basis. 
That is, 
\beq
 \Psi_{N} (\bir_{\rm p},\bir_{\Lambda}) 
 = \sum_M U_{NM} \Phi^{({\rm p\Lambda},J^{\pi})}_M (\bir_{\rm p},\bir_{\Lambda}), 
\eeq
where $H_{3b} \ket{\Psi_N} = E_N \ket{\Psi_N}$ 
with the expansion coefficients, $\{U_M\}$. 
Here $M=\{\kappa,\lambda\}$ is the simplified label for 
the uncorrelated basis. 

In the following, I limit my discussion to only 
with the $J^{\pi}=1^-$ and $2^-$ configurations. 
Thus, the dominant channel is trivially $\kappa(p_{3/2})\otimes \lambda(s_{1/2})$. 
Notice that, in this channel, 
the s.p. proton state is resonant, whereas the $\Lambda$ state is bound.

The expectation value of the three-body Hamiltonian 
gives the total separation energy with respect 
to the $\alpha$-proton-$\Lambda$ breakup threshold. 
That is, 
\beq
  B_{\rm p\Lambda} = -E_{3b} = -\Braket{\Psi | H_{3b} | \Psi}, \label{eq:gero}
\eeq
for an arbitrary state $\ket{\Psi}$. 
In the next section, for the time-evolution calculation 
to describe the 1p emission, Eq. (\ref{eq:gero}) is utilized 
to evaluate the mean three-body energy of the initial state, $\ket{\Psi(t=0)}$. 

\subsection{Experimental data}
The three-body separation energy in Eq. (\ref{eq:gero}) can be 
associated with the two-body separation energies of the subsystems. 
Because of the two different orders to strip the valence proton and $\Lambda$ 
from the $\alpha$ core, one can formulate it in the two ways. 
That is, 
\beqa
 B_{\rm p\Lambda} (^{6}_{\Lambda}{\rm Li}) 
 &=& B_{\rm \Lambda} (^{6}_{\Lambda}{\rm Li})   + B_{\rm p}(^{5}{\rm Li}), ~~{\rm or} \nonumber \\
 &=& B_{\rm p} (^{6}_{\Lambda}{\rm Li})   + B_{\rm \Lambda} (^{5}_{\Lambda}{\rm He}). \label{eq:biry1}
\eeqa
Here $B_{\rm p}(X)$ and $B_{\Lambda}(X)$ indicate 
the single-proton and single-$\Lambda$ 
separation energies from the $X$ nuclide, respectively. 
Note that, if the separation energy is positive (negative), 
that channel is bound (unbound). 
Thus, if $B_{\rm p} (^{6}_{\Lambda}{\rm Li})$ is negative, the 
spontaneous 1p emission can take place. %even when $B_{\rm p\Lambda} (^{6}_{\Lambda}{\rm Li})$ is positive. 
That is, 
\beqa
 B_{\rm p} (^{6}_{\Lambda}{\rm Li}) = 
 B_{\rm \Lambda} (^{6}_{\Lambda}{\rm Li})   + B_{\rm p}(^{5}{\rm Li}) - B_{\rm \Lambda} (^{5}_{\Lambda}{\rm He}). \label{eq:biry2}
\eeqa %Note that $B_{\rm \Lambda}=B_{\Lambda}$, which is the single-Lambda binding energy. 
From the reference data \cite{1975Gal,1981Bertini,1995Noumi,02Till}, 
this is estimated as %$B_{\rm p}(^{6}_{\Lambda}{\rm Li})= -0.6 \pm 0.8$ MeV. 
$B_{\rm p}(^{6}_{\Lambda}{\rm Li}) \simeq -0.6$ MeV, 
and thus, %$B_{\Lambda}(^6_{\Lambda}{\rm Li})=4.5\pm 0.5$ MeV \cite{1981Bertini,1995Noumi}. 
$^6_{\Lambda}$Li may be active for the spontaneous 1p emission 
from its ground state. 
However, it should be also noticed that a large ambiguity 
of the order of $0.1-1$ MeV still remains 
in $B_{\rm p}(^{6}_{\Lambda}{\rm Li})$. 
This ambiguity is from the experimental uncertainties in 
$B_{\rm p}(^{5}{\rm Li})$ \cite{88Ajzen,02Till,NNDCHP} and 
$B_{\Lambda} (^{6}_{\Lambda}{\rm Li})$ \cite{1981Bertini,1995Noumi,1967Harmsen,1972Bha}. 
Concerning these uncertainties, 
it should be still an open question 
whether the 1p emission from the ground state of 
$^{6}_{\Lambda}$Li occurs or not. 

According to the several experimental searches \cite{1967Harmsen,1967Good}, 
the narrow resonance of proton-$^5_{\Lambda}$He has not been reported. 
There can be two interpretations of this result. 
The first one is simply that $^{6}_{\Lambda}$Li is bound 
against the 1p emission. 
The alternative is that 1p-resonance actually exists but 
with a considerably large width. 
In the following discussion, one can find that this problem 
is strongly dependent on the effective proton-$\Lambda$ interaction strength.

\section{Result and Discussion} \label{sec:3}
First I briefly mention the case, where the spontaneous 
1p emission is forbidden. 
In the present model, $B_{\rm \Lambda} (^{5}_{\Lambda}{\rm He})=3.12$ MeV 
as reproduced by the $V_{\alpha-\Lambda}(\bir_{\Lambda})$ consistently 
to the experimental data \cite{1975Gal}. 
Also, the $\alpha$-proton subsystem is {\it unbound} with $B_{\rm p}(^{5}{\rm Li})=-1.69$ 
MeV consistently to Ref. \cite{02Till}. 
From Eqs. (\ref{eq:biry1}) and (\ref{eq:biry2}), when the 1p emission 
is forbidden as $B_{\rm p} (^{6}_{\Lambda}{\rm Li}) >0$, 
it coincides that $E_{3b}=-B_{\rm p\Lambda} <-3.12$ MeV, or equivalently 
that $B_{\Lambda} (^{6}_{\Lambda}{\rm Li}) >4.81$ MeV. 
Remember that $B_{\rm p\Lambda}$ or $B_{\Lambda} (^{6}_{\Lambda}{\rm Li})$ 
is controlled by the interaction strength, $w_0$, which is linked to 
the proton-$\Lambda$ scattering length, $a_v$.

It is confirmed that, if the spontaneous 1p emission is 
forbidden, it requires that $a_v \leq -3.1~(-3.3)$ fm 
for the $J^{\pi}=1^-~(2^-)$ configuration. %in order to reproduce $B_{\rm p} (^{6}_{\Lambda}{\rm Li}) \geq 0$. 
In this case, the three-body ground state is completely bound, and 
the valence proton and $\Lambda$ mostly 
occupy the $(p_{3/2})$ and $(s_{1/2})$ orbits, respectively: 
the occupation probability is obtained as $P(p_{3/2},s_{1/2}) \cong 98~\%$. 
This dominance is attributable to the bound-$(s_{1/2})$ orbit of 
the $\Lambda$ particle.

On the other hand, there have been several realistic YN interaction 
models, which predict the larger scattering length than 
our 1p-binding border, $-3.1$ fm. 
For instance, see Ref. \cite{02Nemura} for a summary of these values. 
In this case, the 1p emission may happen from the ground state of $^6_{\Lambda}$Li. 
In the following, along this second scenario, I discuss the $\Lambda$-inclusion 
effect on the 1p-emission process. 
For this purpose, the time-dependent method is employed.

\subsection{Time-dependent method} \label{subsec:TD}
General formalism of the time-dependent method 
can be found in Refs. \cite{14Oishi,17Oishi}. 
Thus, in the following, I only describe the necessary 
contents for this work.

As discussed in the previous section, the 1p emission is expected 
to occur with the energy release of 
$Q_{\rm 1p}= -B_{\rm p}(^{6}_{\Lambda}{\rm Li}) \simeq +0.6$ MeV. 
To evaluate the three-body energy consistently to the expected 
$Q_{\rm 1p}$ value, it is convenient to reformulate Eq. (\ref{eq:biry1}). 
That is, 
\beq
 Q_{\rm 1p}=-B_{\rm p}(^6_{\Lambda}{\rm Li}) = E_{3b}+B_{\rm \Lambda} (^{5}_{\Lambda}{\rm He}), \label{eq:Q1p}
\eeq
where $B_{\rm \Lambda} (^{5}_{\Lambda}{\rm He})=3.12$ MeV 
in the present model consistently to Ref. \cite{1975Gal}. 
Thus, for fitting to the expected $Q_{\rm 1p}$ value, 
it requires that 
$E_{3b}=\Braket{H_{3b}} \simeq -2.5$ MeV. 
Note also that the $\Lambda$-binding energy 
of $^6_{\Lambda}$Li is given as, 
\beq
 -B_{\Lambda}(^6_{\Lambda}{\rm Li}) = E_{3b} +B_{\rm p}(^5{\rm Li}), \label{eq:BLEV}
\eeq
where $B_{\rm p}(^5{\rm Li})=-1.69$ MeV \cite{02Till}. 
Thus, the expected $Q_{\rm 1p}$ value corresponds to that 
$B_{\Lambda} (^{6}_{\Lambda}{\rm Li}) \simeq +4.2$ MeV.

\begin{figure}[tb] \begin{center}
 \includegraphics[width=0.92\hsize]{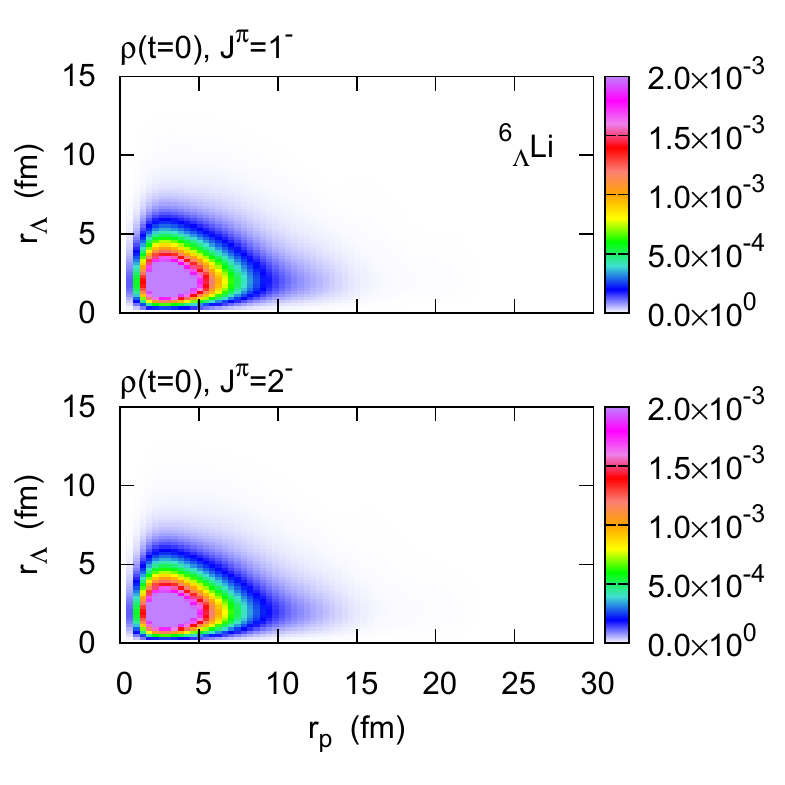} %{./hoge_06_t0_rho.pdf}
 \caption{Probability-density distribution at $t=0$ obtained with the 
confining potential procedure. } \label{fig:mov_01}
\end{center} \end{figure}

In order to fix the initial state for the time-dependent 1p emission, 
which should be associated with the expected $Q_{\rm 1p}$ value, 
I utilize the confining-potential procedure. 
This procedure has provided a good approximation for nuclear 
resonance processes \cite{87Gurv,88Gurv,04Gurv,12Maru,14Oishi,17Oishi}. 
Also, for the two-proton emitter 
$^6$Be \cite{14Oishi,17Oishi}, it has provided the consistent 
result with the non-Hermitian calculation of 
the complex-scaling method \cite{14Myo_Rev,2017Jaga}. 

The confining potential at $t=0$ is given as 
\beq
 V_{\alpha-i}^{(c)}(\bir_i) = V_{\alpha-i}(\bir_i) + \Delta V_{\alpha-i}(\bir_i). 
\eeq %where $i=p$ or $\Lambda$. 
The gap potential $\Delta V_{\alpha-i}(\bir_i)$ is determined 
within the same manner in Ref. \cite{14Oishi}: 
the wall potential at the border radius, $\abs{\bir_i}\geq r_b=5.7$ fm, 
is employed. 
Indeed, it is checked that, as long as the initial-wave packet 
is well confined around the $\alpha$ core, the conclusion 
does not change even with the different $r_b$ values.

Within the confining potential, the initial state can be 
solved by expanding it on the eigen-states of $H_{3b}$. 
That is, 
\beq
 \ket{\Psi(t=0)} = \sum_{N} F_N(0) \ket{\Psi_N}, \label{eq:igvak}
\eeq
where $H_{3b} \ket{\Psi_N} = E_N \ket{\Psi_N}$. 
Figure \ref{fig:mov_01} displays the initial density of probability 
for the $J^{\pi}=1^-$ and $2^-$ configurations. 
That is, 
\beq
 \rho(r_{\rm p},r_{\Lambda}) = 8 \pi^2 r_{\rm p}^2 r_{\Lambda}^2 \int_{-1}^{1} d(\cos \theta_{\rm p\Lambda}) \abs{\Psi(t=0;\bir_{\rm p},\bir_{\Lambda})}^2. 
\eeq
Namely, the original distribution is integrated with respect to 
the proton-$\Lambda$ opening angle, $\theta_{\rm p\Lambda}$ \cite{2005HS,14Oishi}. 
In Fig. \ref{fig:mov_01}, one can find that both the valence 
proton and $\Lambda$ are well confined around the $\alpha$ core. 
In this state, 
the $(p_{3/2},s_{1/2})$ configuration of proton and $\Lambda$ is 
dominant: 
$P(p_{3/2},s_{1/2}) = 97.9~(98.3)~\%$ for $J^{\pi}=1^-~(2^-)$. 
This is similar to the 1p-bound state discussed already. 
Notice also that, due to the different orbits of the ingredient particles, 
the initial density has a long tail with respect to $r_p$, 
whereas it has a compact form with respect to $r_{\Lambda}$. 

The three-body energy is obtained as $E_{3b}=\Braket{H_{3b}}=-2.52$ and $-2.49$ MeV 
for the $J^{\pi}=1^-$ and $2^-$ cases, respectively. 
Here the vacuum-scattering length is fixed as $a_v=-1.6$ fm, 
which corresponds to $w_0=-468.83~\rm MeVfm^3$, 
for the fitting purpose to the expected $Q_{\rm 1p}$ value. 

\begin{figure}[tb] \begin{center}
 \includegraphics[width=0.92\hsize]{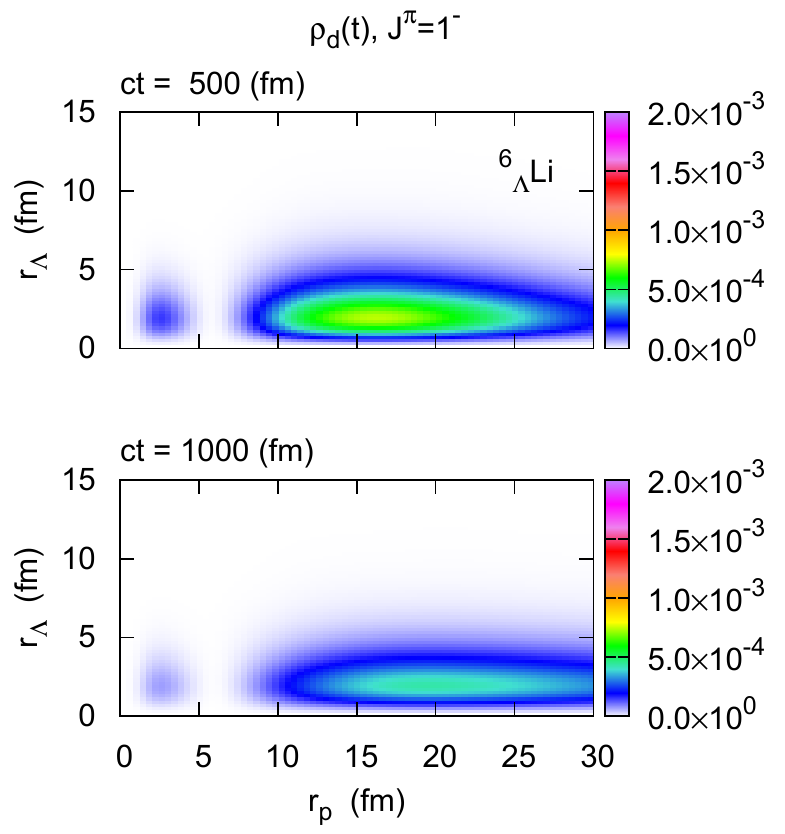}
 \caption{Time-dependent probability-density distribution of the decay state, 
$\rho_d(t)$, for $J^{\pi}=1^-$. } \label{fig:mov_02}
\end{center} \end{figure}

From the initial state in Eq. (\ref{eq:igvak}), 
the time development can be solved as, 
\beq
 \ket{\Psi(t)} = \sum_{N} F_N(t) \ket{\Psi_N}, 
\eeq
where $F_N(t) = e^{-itE_N/\hbar} F_N(0)$. 
It is convenient to define the decay state as, 
\beq
 \ket{\Psi_d(t)} \equiv \ket{\Psi(t)} -\beta(t) \ket{\Psi(0)}, 
\eeq
where $\beta(t)=\Braket{\Psi(0)|\Psi(t)}$ \cite{08Bertulani}. 
Since the initial state is well confined, this decay state represents 
the out-going component from the $\alpha$ core. 

In Fig. \ref{fig:mov_02}, the decay-probability 
density for the $J^{\pi}=1^-$ configuration, 
which is normalized at each point of time, is presented. 
That is, 
\beq
 \rho_d(t;r_{\rm p},r_{\Lambda}) = \frac{8 \pi^2 r_{\rm p}^2 r_{\Lambda}^2}{N_d(t)} \int_{-1}^{1} d(\cos \theta_{\rm p\Lambda}) \abs{\Psi_d(t;\bir_{\rm p},\bir_{\Lambda})}^2, 
\eeq
where $N_d(t)=\Braket{\Psi_d(t)|\Psi_d(t)}$. 
In Fig. \ref{fig:mov_02}, one can clearly see the evacuation of the proton: 
the decay probability grows at $r_p\ge 9$ fm during the time evolution. 
Also, the decay probability is zero for $r_{\Lambda}\ge 6$ fm, indicating 
that the $\Lambda$ particle is always confined around the core. 
This is of course attributable to the bound $s_{1/2}$ orbit. 
Namely, our time-dependent calculation reproduces 
the 1p emission, leaving the bound $\alpha$-$\Lambda$ subsystem. 
The similar pattern of the decay probability is confirmed also in the 
$J^{\pi}=2^-$ case.

\begin{figure}[tb] \begin{center}
 \includegraphics[width=0.82\hsize]{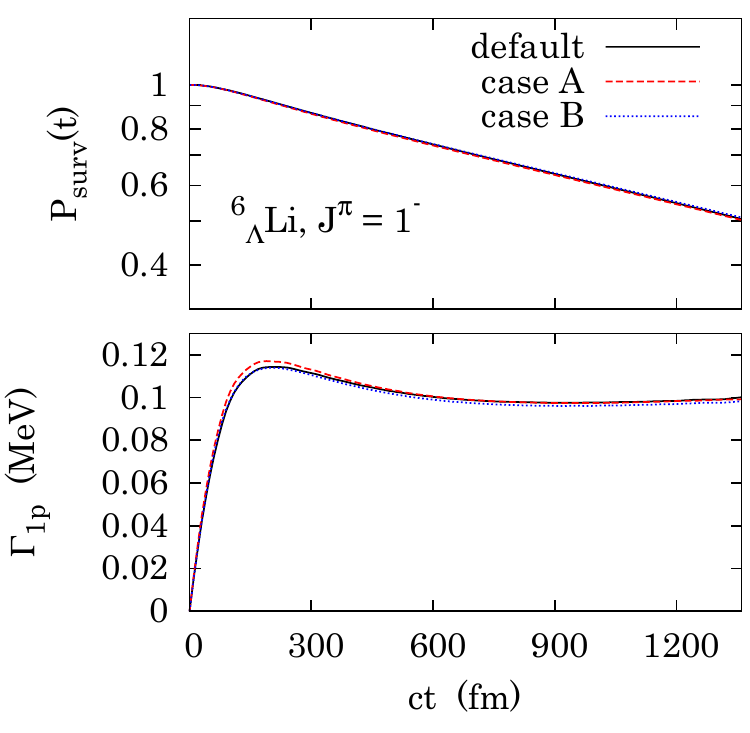}
 \caption{Survival probability and resonance 
width of the 1p emission from the ground state of $^6_{\Lambda}$Li ($J^{\pi}=1^-$). 
The survival probability is plotted in the logarithmic scale. 
In the default case, $R_{\rm box}=120$ fm and $E_{\rm cut}=15$ MeV. 
In the case A, $R_{\rm box}=140$ fm and $E_{\rm cut}=15$ MeV. 
In the case B, $R_{\rm box}=120$ fm and $E_{\rm cut}=20$ MeV. %For other parameters in the model, see the text. 
The mean three-body energy is obtained as $E_{3b}=-2.521$, 
$-2.518$, and $-2.522$ MeV in the default case, case A, and 
case B, respectively. 
} \label{fig:j601}
\end{center} \end{figure}

Figure \ref{fig:j601} shows the survival probability, $P_{surv}(t)$, 
and its decay width, $\Gamma_{\rm 1p}(t)$, for the $J^{\pi}=1^-$ configuration. 
These quantities are defined as the same in Refs. \cite{12Maru,14Oishi}. 
From this result, the process is well interpreted as the 
exponential decay after sufficient time: $\Gamma_{\rm 1p}(ct>600~{\rm fm}) \cong const$. 
The 1p-resonance width is suggested as $\Gamma_{\rm 1p} \cong 0.1$ MeV, 
when $E_{3b}=-2.5$ MeV consistent with the expected 
$Q_{\rm 1p}$ value, $0.6$ MeV.

Before closing this subsection, I show that the conclusion 
does not change even within the larger model space. 
In Fig. \ref{fig:j601}, in addition to the default 
setting with $E_{\rm cut}=15$ MeV and $R_{\rm box}=120$ fm, 
I repeat the same calculation but changing the 
cutoff parameters. 
In the case A, the radial box is extended as $R_{\rm box}=140$ fm, 
whereas the other parameters remain unchanged. 
In the case B, only the cutoff energy is 
increased as $E_{\rm cut}=20$ MeV. 
Consequently, in all the cases, the three-body energy 
and the 1p-emission width show a good coincidence. 
The deviation in the three cases is sufficiently 
smaller than the experimental uncertainties.

\begin{figure}[tb] \begin{center}
 \includegraphics[width=0.82\hsize]{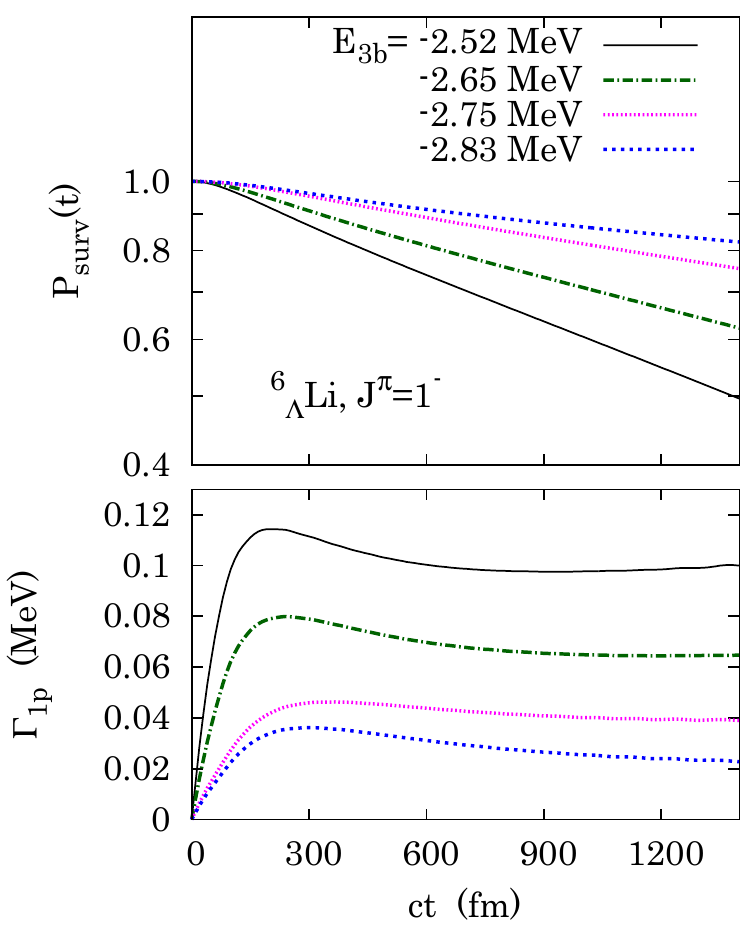}
 \caption{Survival probability and resonance width of the 1p emission 
from the $^6_{\Lambda}$Li nucleus ($J^{\pi}=1^-$). 
Results with $a_v=-1.6$, $-1.8$, $-2.0$, and $-2.2$ fm, 
in correspondence to $E_{\rm 3b}=-2.52$, $-2.65$, $-2.75$, and $-2.83$ MeV, 
respectively, are presented. 
The survival probability is plotted in logarithmic scale. 
} \label{fig:j623}
\end{center} \end{figure}

\begin{figure}[tb] \begin{center}
 \includegraphics[width=0.82\hsize]{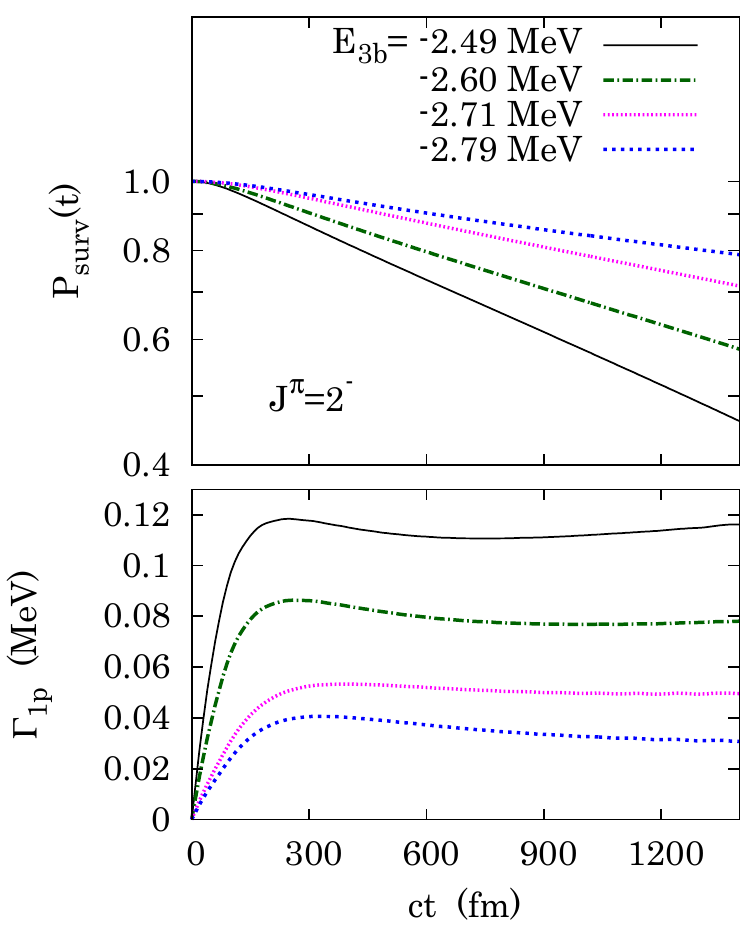}
 \caption{Same to Fig. \ref{fig:j623} but for the 
case of $J^{\pi}=2^-$. 
Results with $a_v=-1.6$, $-1.8$, $-2.0$, and $-2.2$ fm, 
in correspondence to $E_{\rm 3b}=-2.49$, $-2.60$, $-2.71$, and $-2.79$ MeV, 
respectively, are presented. } \label{fig:j624}
\end{center} \end{figure}

\subsection{Sensitivity of 1p-resonance to proton-$\Lambda$ interaction}
In this subsection, the sensitivity of the 1p-resonance energy 
and width is investigated. 
For this purpose, the proton-$\Lambda$ scattering length, $a_v$, is 
treated as the model parameter, which is associated 
with the bare strength of the proton-$\Lambda$ interaction. 

Before going to the detailed discussion, I note the 
range of feasibility of the time-dependent method. 
Fortunately, in this paper, the $Q_{\rm 1p}$ value of interest 
yields the decay width, $\Gamma_{\rm 1p}$, of the order of $0.01-0.1$ MeV. %as shown in Fig. \ref{fig:j263}. 
In this case, after a sufficient time evolution, 
the time-dependent calculation fairly 
reproduces the exponentially decaying rule: 
$P_{surv}(t) \simeq P_{surv}(0) \exp \left(-t\Gamma_{\rm 1p} /\hbar \right)$, where 
$\Gamma_{\rm 1p}$ can be well constant. 
However, I also confirmed that, for the lower $Q_{\rm 1p}$ value, 
the decay width becomes inadequately small, and the time evolution 
should be performed to the long-time scale. 
At present, this calculation is impractical. 
To investigate this long-time scale, the computational improvement 
is necessary: 
the time-dependent method needs to be combined 
with, e.g. the absorption-boundary condition \cite{1998Brinet,2016Schuet}. 
On the other hand, for the larger $Q_{\rm 1p}$ value, 
the decay width becomes so large that it does not 
guarantee the valid picture of the 
quasi-stationary state: the state becomes much alike 
the scattering state, where the continuum effect 
should be more precisely taken into account. 
Also, in such a case, the survival probability does not form the 
exponential decay. 
%Because of the technical difficulty, I leave this task for future work. 

Figures \ref{fig:j623} and \ref{fig:j624} display the results 
of the survival probability and decay width. 
In this result, except $a_v$, the model parameters remain unchanged. 
There, in addition to the previous case, 
I test the new inputs, $a_v=-1.8,-2.0,$ and $-2.2$ fm. %The corresponding interaction strength is $w_0=-502.94$, $-534.02$, and $-562.46~\rm MeVfm^3$, respectively. 
The three-body energy is obtained as $E_{3b}=-2.65,-2.75,$ and $-2.83$ MeV, 
respectively for $J^{\pi}=1^-$, whereas $E_{3b}=-2.60,-2.71,$ and $-2.79$ MeV, 
respectively for $J^{\pi}=2^-$. 
The resultant decay width shows a good convergence after a 
sufficient time evolution, indicating the exponential decay rule. 
These values are in the feasibility range of the time-dependent method. 
It is worthwhile to emphasize that the obtained decay width shows a noticeable fall 
from the 1p-resonance width of the hyperon-less $\alpha$-proton subsystem, 
$\Gamma_{\rm \alpha-p} \simeq 1.1$ MeV \cite{02Till}. 
Namely, the hyperon-inclusion remarkably affects the 
1p-resonance observables. %even though the YN interaction has been suggested to be weaker than the normal nuclear interaction. 

\begin{figure}[tb] \begin{center}
 \includegraphics[width=0.9\hsize]{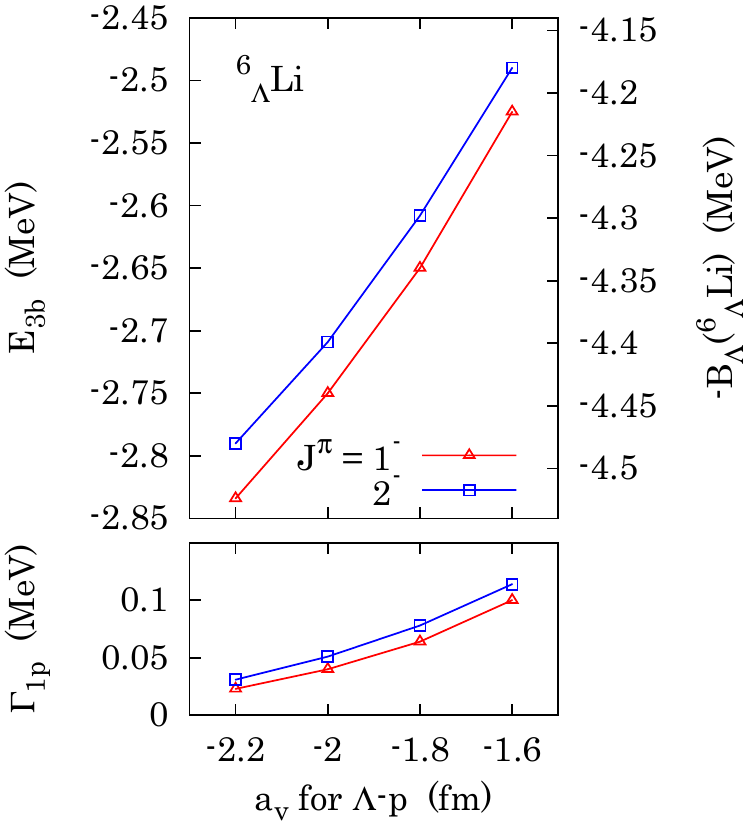}
 \caption{(i) Top panel: three-body energies obtained with several values of 
the input parameter, $a_v$. 
The corresponding $-B_{\Lambda}(^6_{\Lambda}{\rm Li})$ values are 
presented in the right vertical axis according to Eq. (\ref{eq:BLEV}). 
To convert from $E_{3b}$ to the $Q_{\rm 1p}$ value, 
see Eq. (\ref{eq:Q1p}). %$Q=-B_{\rm p}=E_{3b}+B_{\Lambda} (^{5}_{\Lambda}{\rm He})$, where $B_{\Lambda} (^{5}_{\Lambda}{\rm He})=3.12$ MeV \cite{1975Gal}. 
(ii) Bottom panel: 1p-emission width of $^6_{\Lambda}$Li, 
obtained as the mean value during $ct=900-1200$ fm 
in Figs. \ref{fig:j623} and \ref{fig:j624}. } \label{fig:j263}
\end{center} \end{figure}

\begin{figure}[htb] \begin{center}
 \includegraphics[width=0.9\hsize]{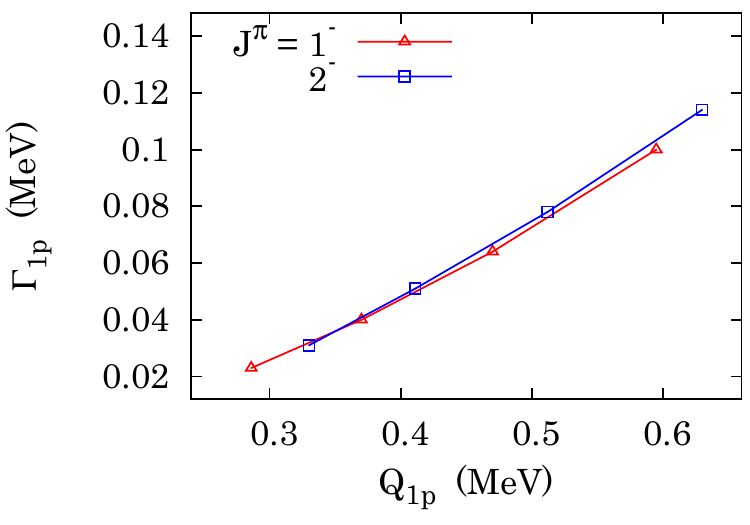}
 \caption{One-proton resonance width as a function of the 
emitted $Q_{\rm 1p}$ value. 
From $E_{3b}$ values in Fig. \ref{fig:j263}, 
Eq. (\ref{eq:Q1p}) is utilized to evaluate $Q_{\rm 1p}$. 
} \label{fig:j478}
\end{center} \end{figure}

In Fig. \ref{fig:j263}, the 1p-resonance observables as functions 
of the vacuum-scattering length, $a_v$, are summarized. %One can clearly see an increase of the resonance energy and width when $a_v$ is increased. 
Note that lowering the $a_v$ value leads to the enhancement of the 
attractive proton-$\Lambda$ force. 
This enhancement naturally yields the decrease 
of the three-body energy as well as the $Q_{\rm 1p}$ value. 
Also, the decrease of the width can be understood from the kinetic effect \cite{2016KM}. 
To show this kinetic effect better, in Fig. \ref{fig:j478}, 
the $Q_{\rm 1p}$-$\Gamma_{\rm 1p}$ relation is presented. 
It is clearly shown that, 
for the lower $Q_{\rm 1p}$ value, $\Gamma_{\rm 1p}$ becomes smaller, 
since the 1p penetrability via the core-proton 
potential barrier is more suppressed. 
Consequently, with the stronger attraction between proton-$\Lambda$, 
the system becomes more long-lived.

\subsection{Spin-dependent splitting} \label{subsec:sd_split}
In the previous subsection, the same parameter is used 
for the spin-singlet and spin-triplet proton-$\Lambda$ interactions. 
As the result, the splitting energy between the $1^-$ and 
$2^-$ resonances is much smaller than the expected 
value from the standard, spin-dependent YN interaction. 

In this subsection, I extend the model to take 
the spin-dependence of the proton-$\Lambda$ interaction into account. 
For this purpose, instead of Eq. (\ref{eq:yutwo}), 
I employ a spin-dependent contact interaction: 
\beq
 v_{\rm p-\Lambda}(\bir_{\rm p},\bir_{\Lambda}) 
 = \delta(\bir_{\rm p}-\bir_{\Lambda}) \left[ w^{(t)}_0 \hat{P}_{t} 
   + w^{(s)}_0 \hat{P}_{s} \right], 
\eeq
where $\hat{P}_{s~(t)}$ indicates the projection to the spin-singlet 
(triplet) proton-$\Lambda$ configuration. 
Its strength parameters, $w^{(s)}_0$ and $w^{(t)}_0$, are determined 
from the spin-singlet and triplet scattering lengths, respectively. 
That is, 
\beq
 w^{(x)}_0 = \frac{2\pi^2 \hbar^2 a^{(x)}_v}{\mu_{\rm p-\Lambda} (\pi-2a^{(x)}_v k_{\rm cut})} 
~~~~({\rm MeVfm^3}), 
\eeq
where $x=s$ or $t$. 

For the input parameters, $(a^{(s)}_v,a^{(t)}_v)$, two cases as 
tabulated in TABLE \ref{table:vuyds} are compared: 
in the cases I and II, for the spin-singlet part, 
$a^{(s)}_v=-2.2$ fm and $-2.5$ fm, respectively. 
On the other hand, for the spin-triplet part, it 
is fixed as $a^{(t)}_v=-1.7$ fm in both cases. 
Namely, I consider the small (large) difference between the 
spin-singlet and triplet forces in the case I (II). 
These parameters are chosen consistently to the empirical 
results predicted by several YN interactions \cite{2016Gal,02Nemura}. 
Other parameters in the model remain unchanged.

\begin{table}[b] \begin{center}
\caption{Input parameters of the vacuum-scattering lengths in 
the spin-singlet and triplet channels for the 
proton-$\Lambda$ contact interaction. 
The 1p-energy release, 
$Q_{\rm 1p}=-B_{\rm p}(^6_{\Lambda}{\rm Li})$, is evaluated 
from Eq. (\ref{eq:Q1p}). } \label{table:vuyds}
  \catcode`? = \active \def?{\phantom{0}} %define `?' as ' '(one-blank).
  \begingroup \renewcommand{\arraystretch}{1.2}
  \begin{tabular*}{\hsize} { @{\extracolsep{\fill}} ccccc} \hline \hline
           &$a^{(s)}_v$ (fm)   &$a^{(t)}_v$ (fm)   &\multicolumn{2}{c}{$Q_{\rm 1p}$ (MeV)}  \\ %\cline{4-5}
           &              &           &$J^{\pi}=1^-$    &$J^{\pi}=2^-$ \\ \hline
  ~case I??  &$-2.2$     &$-1.7$      &$+0.29$         &$+0.53$   \\
  ~case II?  &$-2.5$     &(same)      &$+0.22$         &$+0.53$   \\
  ~case III  &$\le -4.0$ &(same)      &(bound)         &$+0.53$   \\ \hline \hline
  \end{tabular*}
  \endgroup
  \catcode`? = 12 %initialize `?'.
\end{center} \end{table}

In TABLE \ref{table:vuyds}, the resultant $Q_{\rm 1p}$ 
values are presented in the cases I and II. 
The mean three-body energy, $E_{3b}$, is obtained with the 
initial state, which is solved within the same confining potential 
in Sec. \ref{subsec:TD}. 
Consequently, in contrast to the previous results, 
now there is a noticeable splitting of the $1^-$ and $2^-$ 1p-resonance 
energies due to the spin-dependent proton-$\Lambda$ force. 
In the $J^{\pi}=1^-$ configuration, a noticeable 
sensitivity of the $Q_{\rm 1p}$ value (resonance energy) to the spin-singlet force 
is found: when $a^{(s)}_v$ becomes small, the $E_{3b}$ as well as $Q_{\rm 1p}$ 
gets decreased, consistently with that the spin-singlet 
proton-$\Lambda$ interaction, $w^{(s)}_0$, becoming more attractive. 
On the other hand, the result of the $J^{\pi}=2^-$ configuration 
is almost independent of the changes of the spin-singlet force. 
This can be understood from that, in the $(\kappa,\lambda)=(p_{3/2},s_{1/2})$ channel 
of the proton-$\Lambda$, only the spin-triplet configuration is 
allowed to couple to $J^{\pi}=2^-$. 
Indeed, this channel is dominant with $P(p_{3/2},s_{1/2})\simeq 98$\% 
in both the cases I and II. 
%Thus, the $2^-$ three-body energy is insensitive to the spin-singlet scattering length. 

\begin{figure}[tb] \begin{center}
 \includegraphics[width=0.98\hsize]{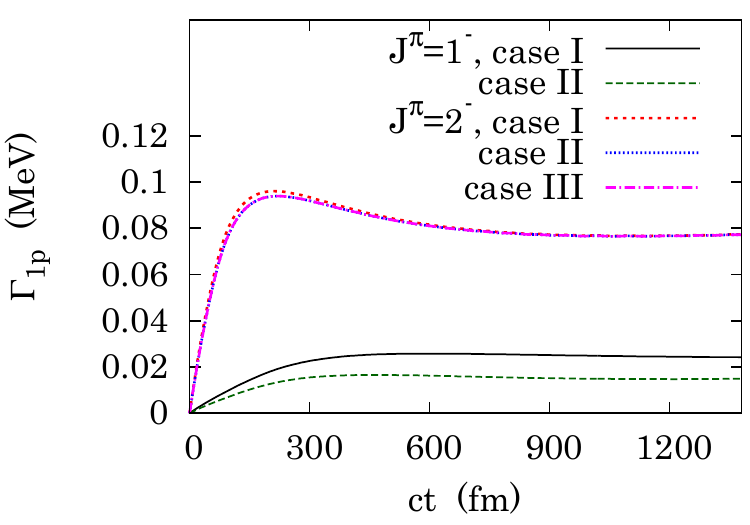}
 \caption{The 1p-emission width of $^{6}_{\Lambda}$Li. 
These results are obtained with the sets of parameters, 
$(a^{(s)}_v,a^{(t)}_v)$, tabulated in TABLE \ref{table:vuyds}. 
In the case III, 1p emission is forbidden 
from the $1^-$ ground state. } \label{fig:08_SD}
\end{center} \end{figure}

In Fig. \ref{fig:08_SD}, the 1p-emission width, $\Gamma_{\rm 1p}(t)$, 
from the time-dependent calculation is presented. 
First, in the $2^-$ configuration, as expected from the insensitivity 
of the $Q_{\rm 1p}$ value, the width is almost unchanged in the 
cases I and II: $\Gamma_{\rm 1p} \cong 75$ keV after the 
convergence at $ct \ge 600$ fm. 
In the $1^-$ configuration, on the other hand, 
$\Gamma_{\rm 1p} \cong 25$ and $15$ keV in the 
cases I and II, respectively. 
This sensitivity can be understood from the kinetic effect 
with the different $Q_{\rm 1p}$ values, as discussed in the 
last subsection. 
Notice that these predicted widths are comparably smaller than 
the splitting energy, $Q_{\rm 1p}(2^-)-Q_{\rm 1p}(1^-) \cong 200$ - $300$ keV, 
between the $1^-$ and $2^-$ states. 
Thus, there can be the two well-separated 1p-resonance levels 
in $^6_{\Lambda}$Li, which may provide suitable data to optimize the 
theoretical YN-interaction model with the spin-dependence.

Before closing the discussion, it is worth mentioning the last possible case, 
where the 1p emission is forbidden from the $1^-$ state, whereas 
it is still allowed from the $2^-$ state. 
For this condition, it requires $a^{(s)}_v \le -4.0$ fm, 
whereas $a^{(t)}_v$ is kept as $-1.7$ fm. 
This set is listed as the case III in TABLE \ref{table:vuyds}. 
In this case, as the result, $Q_{\rm 1p}<0$ for $J^{\pi}=1^-$, 
meaning that the 1p emission is not allowed. 
Namely, this spin-singlet force is sufficiently attractive 
to bind the valence proton. %corresponding to the low $a^{(s)}_v$ value. 
Consequently, if the spin-dependent splitting is sufficiently wide, 
the $1^-$ ground state of $^6_{\Lambda}$Li may be 
stable against the 1p emission. 
Note also that, even in this case, the 1p emission from the $2^-$ state 
can be still active, because of the insensitivity of the $Q_{\rm 1p}$ energy. 
Its width is also insensitive to $a^{(s)}_v$ as shown in Fig. \ref{fig:08_SD}.

%<><><><><><><><><><> CORRECTED 2nd <><><><><><><><><><><><><><<><><><><>

\section{Summary} \label{sec:4}
It is shown that the spontaneous-1p emission from $^6_{\Lambda}$Li can be sensitive 
to the proton-$\Lambda$ scattering length or its corresponding interaction strength. 
This sensitivity can be understood from the kinetic effect 
on the 1p-tunneling process via the core-proton potential. 
The proton-$\Lambda$ interaction, which is the novel ingredient 
in this hypernuclear system, can noticeably affect the 
1p-emission observables, including the resonance energy and width, 
compared with the normal $\alpha$-proton resonance. 
By taking the spin-dependence of the proton-$\Lambda$ scattering length 
into account, a remarkable splitting of the $1^-$ and $2^-$ 
resonances is suggested. 
Utilizing the empirical spin-singlet and spin-triplet proton-$\Lambda$ 
scattering lengths as the model parameters, 
the decay widths of the two resonances are evaluated. 
From these suggestions, one may expect that the simultaneous 
1p emission can be a suitable tool to investigate 
the YN-interaction properties.

It is worth mentioning that, 
from the resultant $\Gamma_{\rm 1p}$ value, 
the 1p emission from $^6_{\Lambda}$Li is suggested to occur in the 
timescale of $\tau=\hbar/\Gamma_{\rm 1p} \simeq 10^{-20}-10^{-21}$ seconds. 
This timescale is sufficiently shorter than the 
self-decay lifetime of $\Lambda$, $10^{-10}$ seconds. 
Thus, one can infer that the $\Lambda$ particle, as well as the 
daughter nucleus, $^5_{\Lambda}$He, is comparably unbroken 
during the emission process. 
This speculation, however, turns to be invalid in one exceptional 
case, where the proton-$\Lambda$ attraction is strong enough 
to prohibit the 1p emission.

In this paper, I utilized the simple three-body model. 
Compared with the ab initio multi-nucleon analysis \cite{02Nemura,2003Usm,2016Wirth,2016Gazda}, 
this model includes several schematic parts. 
In forthcoming studies, I intend to combine the time-dependent 
method with the ab initio calculation, 
in order to investigate the particle-bound and unbound hypernuclei 
on equal footing. 
There are several topics, which can be addressed 
after this development. 
Those include, e.g., 
three-body hypernuclear interaction \cite{02Nemura,2003Usm,2016Wirth}, 
spin-dependent charge-symmetry breaking \cite{2016Gazda,2015Yama}, and 
tensor force effect \cite{04Ukai}. 
How these effects are reflected on the nucleon-resonance properties 
can be a natural question following this first study. 
Also, the time-dependent calculation with the standard 
YN interaction model is an important future work.

For the first attempt to discuss the 1p emission with strangeness, 
the lightest hyper-1p emitter is investigated in this paper. 
In future studies, the similar interest can be widely devoted to 
the other systems along the proton-drip and neutron-drip lines, 
as well as the excited states of hypernuclei. 
For these systems, the structural model may need 
to be expanded: one has to implement the time-dependence 
into the four-or-more-particle model, or the fully 
microscopic-structure model. 
Several developments along this direction are in progress now. 
%Our time-dependent method, which will be combined with these 
%adequate structural models, can be a suitable tool 
%for this future study. 

\begin{acknowledgments}
T. O. sincerely thanks T. O. Yamamoto, L. Fortunato, A. Vitturi, T. Saito, 
A. Pastore, M. Kortelainen, and E. Hiyama for fruitful discussions. 
T. O. acknowledges the financial support within 
the P.R.A.T. 2015 project {\it IN:Theory} of the University 
of Padova (Project Code: CPDA154713). 
The computing facilities offered by CloudVeneto (CSIA Padova and I.N.F.N. Sezione di Padova) 
are acknowledged. 
%We also acknowledge the CSC-IT Center for Science Ltd., Finland, 
%for the allocation of computational resources. 
%This work was supported by Academy of Finland and University of 
%Jyv\"{a}skyl\"{a} within the FIDIPRO programme. 
\end{acknowledgments}

\appendix*

\section{Transformation of coordinates}
In the main text, the core-orbital coordinates are 
employed for the three-body system. 
In this Appendix, I give a basic formalism to 
determine these coordinates. 
It starts with the original coordinates and the conjugate momenta: 
\beq
  \vec{X} \equiv 
  \left[ \begin{array}{c} 
  \bi{x}_{\rm p} \\ \bi{x}_{\Lambda} \\ \bi{x}_{\rm c} \end{array} \right], \qquad 
  \vec{Q} \equiv 
  \left[ \begin{array}{c} 
  \bi{q}_{\rm p} \\ \bi{q}_{\Lambda} \\ \bi{q}_{\rm c} \end{array} \right]. 
\eeq
With these coordinates, the three-body Hamiltonian is trivially given: 
\beq
 H_{3b} = \sum_{i} \frac{\bi{q}_i^2}{2m_i}
 + V_{\rm c-p} + V_{\rm c-\Lambda} + v_{\rm p-\Lambda}, 
\eeq
where $i={\rm p,\Lambda}$ and c for the valence 
proton, $\Lambda$, and the core, respectively.

With a $3\times 3$ matrix $U$, the core-orbital 
coordinates can be determined in the matrix form: 
\beq
  \vec{R} \equiv 
  \left[ \begin{array}{c} 
  \bir_{\rm p} \\ \bir_{\Lambda} \\ \bir_G \end{array} \right] 
  = U \vec{X}, \quad
  \vec{P} \equiv 
  \left[ \begin{array}{c} 
  \bip_{\rm p} \\ \bip_{\Lambda} \\ \bip_G \end{array} \right] 
  = ({}^{t}U)^{-1} \vec{Q}, 
\eeq
where 
\beq
 U = 
 \left( \begin{array}{ccc} 
    1 & 0 & -1 \\
    0 & 1 & -1 \\
    \frac{m_{\rm p}}{M} & \frac{m_{\Lambda}}{M} & \frac{m_{\rm c}}{M}
 \end{array} \right), 
\eeq
with $M\equiv \sum_i m_i$ (total mass). 
In these new coordinates, the Hamiltonian reads 
\beq
 H_{3b} = \frac{p_G^2}{2M} + \frac{p_{\rm p}^2}{2\mu_{\rm p}} + \frac{p_{\Lambda}^2}{2\mu_{\Lambda}} 
 + \frac{\bip_{\rm p} \cdot \bip_{\Lambda}}{m_{\rm c}} + (potentials), 
\eeq
where the relative masses are given as $\mu_i=m_i m_{\rm c}/(m_i + m_{\rm c})$ 
for $i={\rm p}$ and $\Lambda$. 
The first term represents the center-of-mass kinetic energy. 
Since this center-of-mass term is well separated, 
one can obtain the three-body Hamiltonian in Eq. (\ref{eq:fevag}) correctly. 
Notice that the recoil term, $(\bip_{\rm p} \cdot \bip_{\Lambda})/m_{\rm c}$, 
should be always diagonalized, even if $v_{\rm p-\Lambda}$ is zero.

         %\bibitem{03Nemura} H. Nemura et al., XXX (2099).

\end{document}